\documentclass[letterpaper, 10 pt, conference]{ieeeconf}  

\IEEEoverridecommandlockouts                              
\usepackage[utf8]{inputenc}
\usepackage[T1]{fontenc}
\usepackage{graphicx}
\usepackage{array}
\usepackage{tabularx}
\usepackage{caption}
\usepackage{subcaption}
\usepackage{graphics}

\title{\LARGE \bf
Human Emotions Analysis and Recognition Using EEG Signals in Response to \textbf{360}° Videos}

\author{Haseeb ur Rahman Abbasi$^{1}$,  Zeeshan Rashid $^{1}$, Muhammad Majid $^{1}$, and Syed Muhammad Anwar$^{2}$
\thanks{*This work was not supported by any organization}
\thanks{$^{1}$Haseeb ur Rahman Abbasi, Zeeshan Rashid and Muhammad Majid are with Department of Computer Engineering, University of Engineering and Technology, Taxila, Pakistan.}%
\thanks{$^{2}$Syed Muhammad Anwar is with with Sheikh Zayed Institute for Pediatric Surgical Innovation, Children’s National Hospital, Washington, DC and School of Medicine and Health Sciences, George Washington University, Washington, DC.}%
}

\begin{document}

\maketitle

\begin{abstract}

Emotion recognition (ER) technology is an integral part for developing innovative applications such as drowsiness detection and health monitoring that plays a pivotal role in contemporary society. This study delves into ER using electroencephalography (EEG), within immersive virtual reality (VR) environments. There are four main stages in our proposed methodology including data acquisition, pre-processing, feature extraction, and emotion classification. Acknowledging the limitations of existing 2D datasets, we introduce a groundbreaking 3D VR dataset to elevate the precision of emotion elicitation. Leveraging the Interaxon Muse headband for EEG recording and Oculus Quest 2 for VR stimuli, we meticulously recorded data from 40 participants, prioritizing subjects without reported mental illnesses. Pre-processing entails rigorous cleaning, uniform truncation, and the application of a Savitzky-Golay filter to the EEG data. Feature extraction encompasses a comprehensive analysis of metrics such as power spectral density, correlation, rational and divisional asymmetry, and power spectrum. To ensure the robustness of our model, we employed a 10-fold cross-validation, revealing an average validation accuracy of 85.54\%, with a noteworthy maximum accuracy of 90.20\% in the best fold. Subsequently, the trained model demonstrated a commendable test accuracy of 82.03\%, promising favorable outcomes.
\end{abstract}

\section{INTRODUCTION}
Emotion recognition, a technology with diverse applications, is currently employed in driving drowsiness detection, workload evaluation, and health monitoring, significantly impacting society. It involves both emotion elicitation and classification. Emotion recognition not only portrays behavioral and mental states but also strengthens human-computer interaction. Positive emotions indicate a healthy state, contrasting with negative emotions which could even be linked to conditions such as depression and an elevated suicide risk [1]. Recent studies favor physiological signals like electroencephalography (EEG) over visible signs (like speech, and facial expressions), considering them more accurate indicators of genuine emotions due to their connection to the central nervous system. There has been a significant interest in EEG based analysis,  due to its potential to offer a straightforward, cost-effective, portable, and user-friendly solution for emotion identification [2]. Therefore, EEG finds extensive application in diverse biomedical contexts, playing a crucial role in diverse tasks such as stress assessment [3], the detection of depression disorders [4], and the exploration of schizophrenia [5] to  name a few.

Virtual reality (VR) is a technology that uses computer-generated simulated environments, that give users the illusion of real physical exposure. VR is widely used in different fields like education, medicine, entertainment, defense, marketing, real estate, and many more [6]. Environments that are difficult to realize can be simulated using VR. The 2021 launch of Horizon Worlds by Meta Platforms has ignited debates on the societal implications of the Metaverse, described as "the layer between you and reality." Meta envisions seamlessly integrating avatars and holograms in a 3D virtual shared world for work and social interactions [7]. Although VR technology is making progress rapidly, it has not yet reached the peak of its development, and its limitations cannot be reliably predicted  for now.

At present, most emotion recognition systems are data-dependent. Many datasets have been proposed in the literature for research. Prominent publicly accessible datasets for emotion recognition, such as DEAP [8], DREAMER [9], and ASCERTAIN [10], incorporate EEG signals in conjunction with other physiological data. These datasets used videos, music, and images to elicit different emotions. The main limitation of these datasets is that all these use stimuli that are 2D, non-immersive, and hence they lack the feel of presence for users when they interact with them. So, a 3D virtual environment is required for more effective emotion elicitation.

\begin{figure*}[t]
    \centering
    \includegraphics[width=160mm]{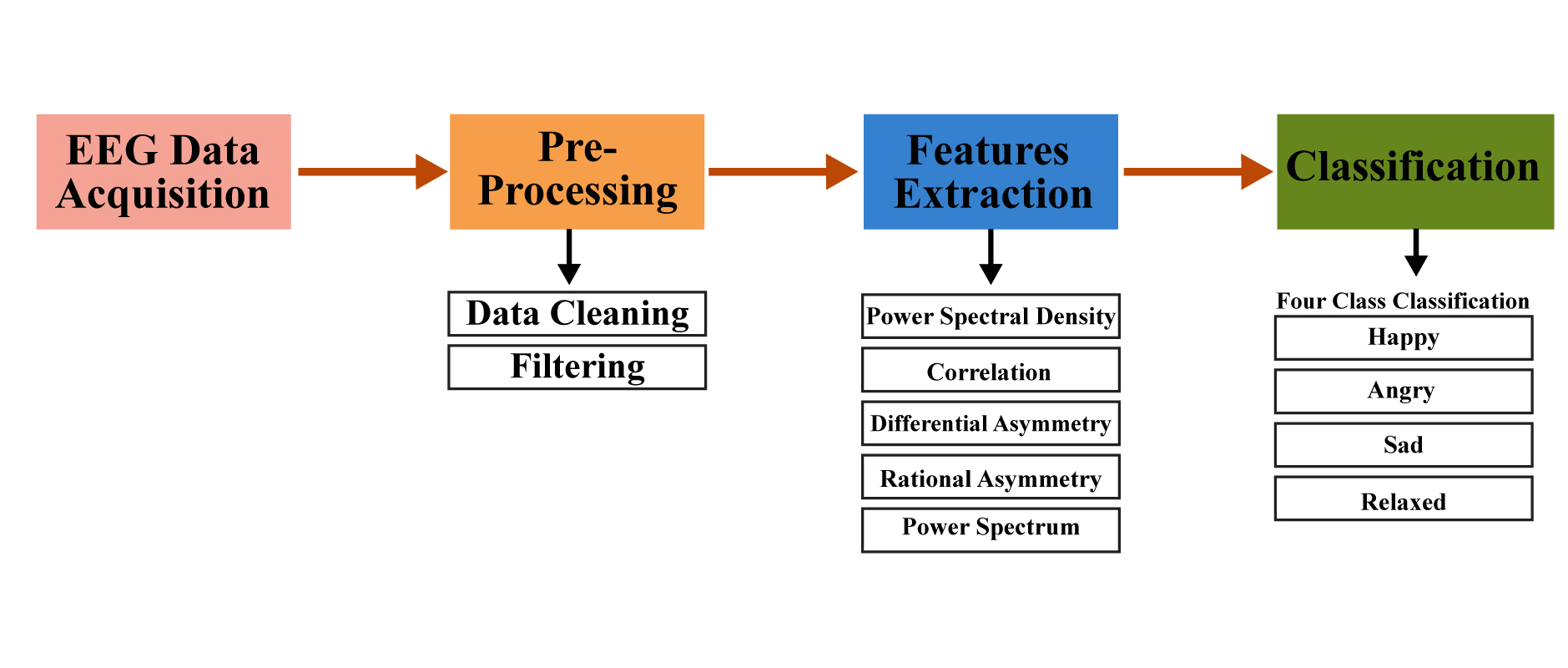}
    \caption{The proposed method for emotion recognition in response to 360° videos}
    \label{fig:1}
\end{figure*}

Studies have shown that emotions elicited using immersive virtual environments are better than those induced using non-immersive methods [11].  Aelee Kim et al [12]  showed that VR gives a more immersive experience and greater emotional response to horror films. The effectiveness of VR as emotion elicitation stimuli has been proven [13-16]. In [17], a VR-based EEG dataset named as VREED,  was presented and used to recognize different emotions. The dataset is currently in the embargo phase and is not yet available publicly. In [18], emotion recognition based on VR was conducted to classify four classes of emotions. Another study [19] was conducted for emotion recognition in response to VR using data from frontal EEG electrodes. 

Due to limited research in emotion recognition in response to VR and the unavailability of a physiological signals dataset, this paper acquires a new dataset for emotion analysis using EEG signals in response to VR environments. In the near future, interactions with VR technology is expected to increase manifold, as leading IT industries have already started developing VR-based applications. So, our work is a significant step forward to minimize the gap towards human emotion analysis and VR interactions. The main contributions of this paper are,

\begin{enumerate}
    \item We curate a new dataset for emotion recognition using EEG signals acquired in response to VR content i.e., 360° videos.
    \item We further perform emotion recognition utilizing variations in EEG signals in response to 360° video using frequency domain features.
\end{enumerate}

The remaining part of the paper is organized as follows. Section II explains the proposed methodology for this study. Section III presents results and discussion and a conclusion is drawn in section IV.

\section{OUR PROPOSED METHODOLOGY}
An overflow workflow for the proposed strategy for human emotion analysis and recognition in response to a virtual reality environment (360° videos) is shown in Figure 1. The four main steps are: 1)data acquisition, 2) pre-processing, 3) feature extraction, and 4) emotion classification. 

\subsection{Data Acquisition}

\textbf{1) Participants and Apparatus:} Data were recorded from a total of 40 subjects (24 males and 16 females). The age range was between 18 to 35 years.  Subjects did not report any mental illness before recording. For EEG recording, we employed the Interaxon Muse headband, a versatile and user-friendly EEG recording system with four channels located at AF7, AF8, TP9, and TP10 positions. The Muse headband records EEG data at 256 Hz and connects to a smartphone via Bluetooth for data transmission. To present the stimuli to subjects, a head-mounted display (HMD) from Meta named Oculus Quest 2  with six degrees of freedom with realistic precision, $1832 \times 1920$ resolution per eye, and with 90 Hz refresh rate was used. 

\textbf{2) Stimuli Selection:} Sixteen videos, four from each quadrant were chosen from a publicly available immersive VR video database, each accompanied by valence and arousal ratings obtained through the Self-Assessment Manikin (SAM) scale [19]. The videos whose ratings were farthest from the origin were selected. From the first quadrant of the valence arousal scale, we selected video numbers 50, 62, 69, and 52. Video numbers 33, 32, 27, and 22 were selected from the second quadrant. For the third quadrant, we selected 15, 3, 1, and 14, and for the fourth quadrant 20, 65, 68, and 21 were selected from [19].  Each quadrant represents happy, angry, sad, and relaxed emotions respectively. These selected videos were then organized into four sessions, each comprising of four videos. The design of each session was aimed to maintain a total duration of approximately 15 minutes, considering user fatigue, as studies suggest discomfort after prolonged viewing periods [19].

\textbf{3) Experimental Procedure:} For EEG recordings, participants were taken to temperature-regulated room with consistent lighting conditions. In this setting, an overview of the experimental process was presented, and participants were invited to sign the consent form. Furthermore, individuals were requested to complete a demographic questionnaire to supply details about their age, gender, and any pertinent information related to mental health. Participants were informed that they had the freedom to quit the experiment at any time of their choosing. The experiments for this study are designed following the Helsinki Declaration and the study was approved by the Board of Advanced Studies Research and Technological Development at the University of Engineering and Technology, Taxila.

\subsection{Data Pre-Processing}

The recorded data was transferred to a computer for pre-processing. Initially, missing data points from raw EEG channels were generated using window-based averaging to ensure data cleanliness. Although, the recorded data had very few of these points. Subsequently, the cleaned data files were uniformly truncated to match the duration of each video. Following this, a Savitzky-Golay filter with a third-order polynomial and a window size of 11 was applied to smooth out potential outliers in the data. Furthermore, five frequency bands: delta (0–4 Hz), theta (4–7 Hz), alpha (8–12 Hz), beta (12–30 Hz), and gamma (30–50 Hz) were obtained from raw EEG channels data.

\subsection{Features Extraction}

To analyze the recorded EEG data, distinct frequency domain feature groups were derived from each channel. These included power spectral density (PSD), correlation (C), divisional asymmetry (DASM), rational asymmetry (RASM), and power spectrum (PS). In particular, PSD characterizes the power distribution across specific frequency ranges. Features from this group included the mean and variance of the PSD for each channel. This feature group comprised 8 features. Correlation, a statistical measure reflecting the degree of variation between two values, was computed for asymmetric channels of the left and right hemispheres—specifically, (TP9, TP10) and (AF7, AF8). Two features were computed for this feature group. DASM represented the variance in absolute power between asymmetric channels of the brain hemispheres, while RASM denoted the ratio of absolute power between left and right hemisphere channels. Four features were extracted from these feature groups. The power spectrum involved the average absolute power across four scalp electrodes in the five frequency bands of the EEG signal. 20 Features were extracted from this group, comprising a total feature vector length of 34 features.

\begin{table}[t]
    \caption{Performance comparison of different SVM kernels for four class classification.}
    \label{table_example}
    \centering
    \begin{tabular}{|>{\centering\arraybackslash}p{1.75cm}|>{\centering\arraybackslash}p{1.75cm}|>{\centering\arraybackslash}p{1.75cm}|>{\centering\arraybackslash}p{1.65cm}|}
        \hline
        Kernel Function & Average accuracy across 10 folds & Maximum Accuracy on the best fold & Test Accuracy on Best Fold \\
        \hline 
        RBF & 70.49\% & 76.92\% & 64.84\% \\
        \hline
        Linear & 72.82\% & 82.69\% & 75\% \\
        \hline
        Gaussian & 69.52\% & 75\% & 64.84\% \\
        \hline
        Polynomial & \textbf{85.54\%} & \textbf{90.20\%} & \textbf{82.03\%} \\
        \hline
    \end{tabular}
\end{table}

\subsection{Classification}

For classification, we used a machine learning algorithm named Support Vector Machine (SVM). In SVM, data items are positioned in an n-dimensional space. The classification of these data points involves the identification of a hyperplane, strategically separating the classes for optimal distinction. SVM employs an iterative training algorithm to pinpoint the most advantageous hyperplane, working towards the minimization of an error function. Using SVM, we classified four classes of emotions, based on videos selected from four quadrants. We applied different kernel functions like radial basis function (RBF), Gaussian, linear, and polynomial to compare results. 

\section{RESULTS AND DISCUSSION}

To train the machine learning model, we split the feature vector into 80 percent training and 20 percent testing sets. 10-fold cross-validation was applied to the training set. We then tested the test set on the model trained with the best fold (with maximum accuracy). SVM was trained using RBF, linear, Gaussian, and polynomial kernels. Using a polynomial kernel we got an average cross-validation accuracy of 85.54\% with a maximum accuracy of 90.20\% on the best fold. On testing the best model with the test set, we achieved a remarkable accuracy of 82.03\%. Table I compares the results of different kernels for four class classifications. The 2nd-degree polynomial kernel outperforms all other kernels.

\begin{figure}
    \centering
    \begin{subfigure}{0.23\textwidth}
        \includegraphics[width=\linewidth]{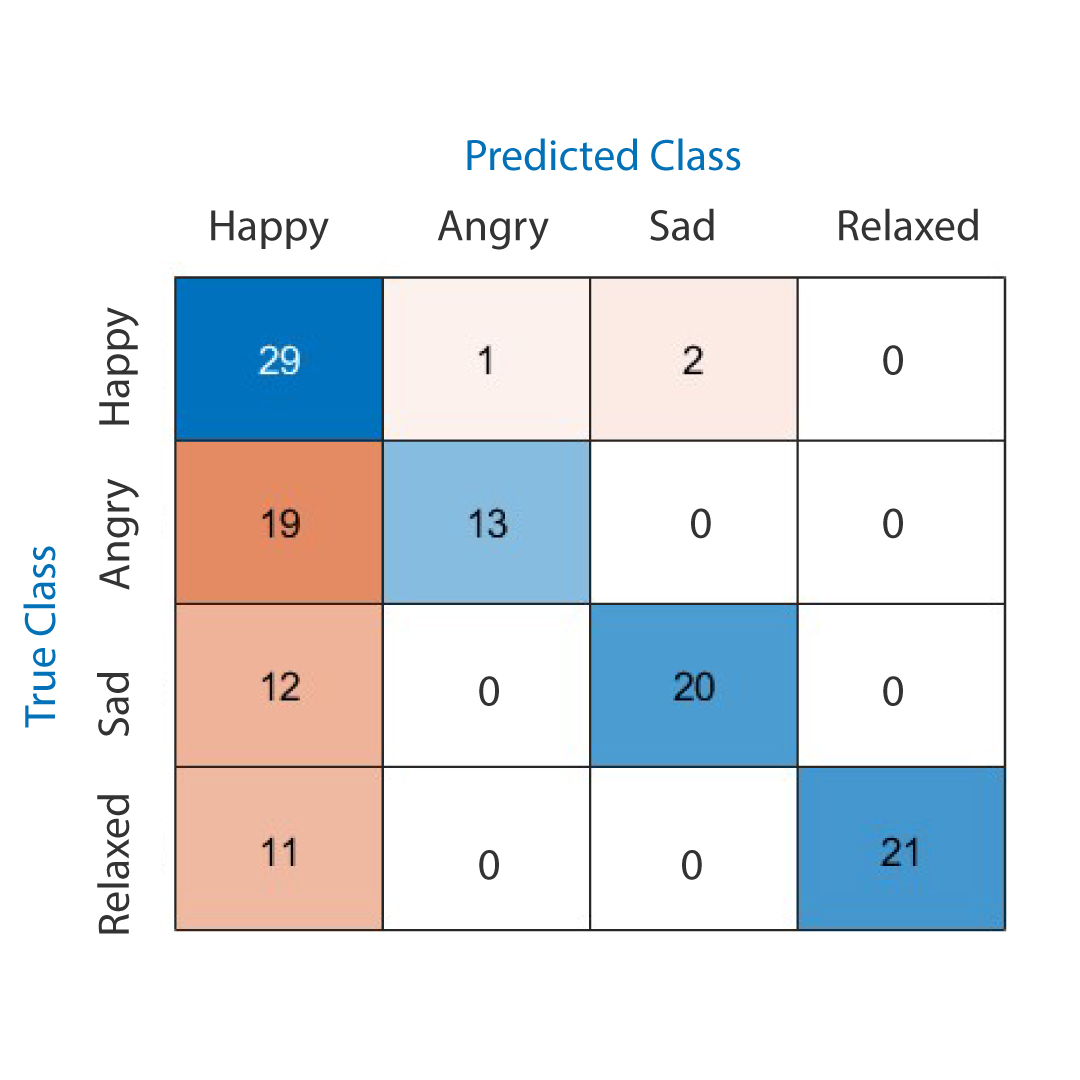}
        \caption{}
        \label{fig:subfig_a}
    \end{subfigure}
    \hfill
    \begin{subfigure}{0.23\textwidth}
        \includegraphics[width=\linewidth]{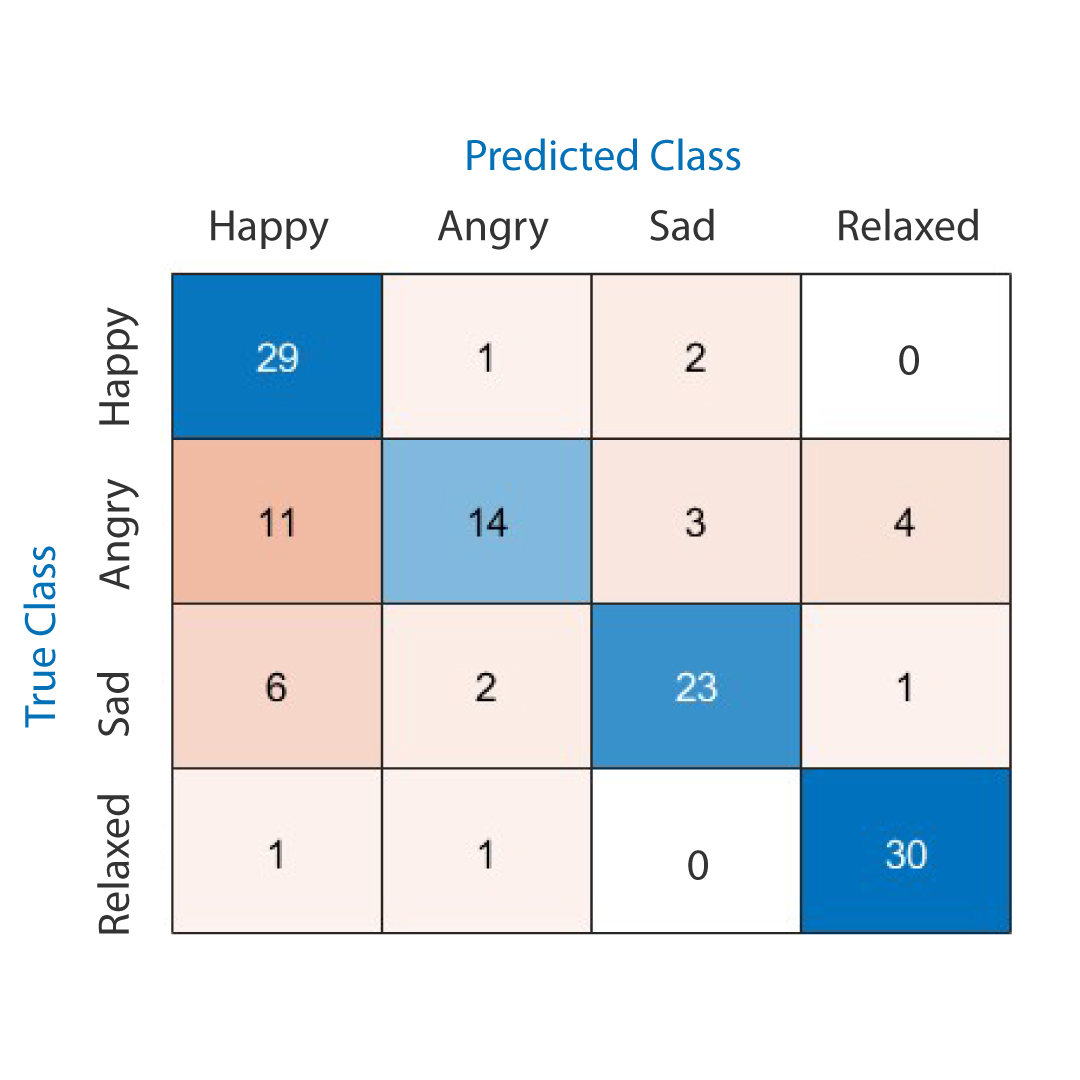}
        \caption{}
        \label{fig:subfig_b}
    \end{subfigure}

    \vspace{1em} 
    
    \begin{subfigure}{0.23\textwidth}
        \includegraphics[width=\linewidth]{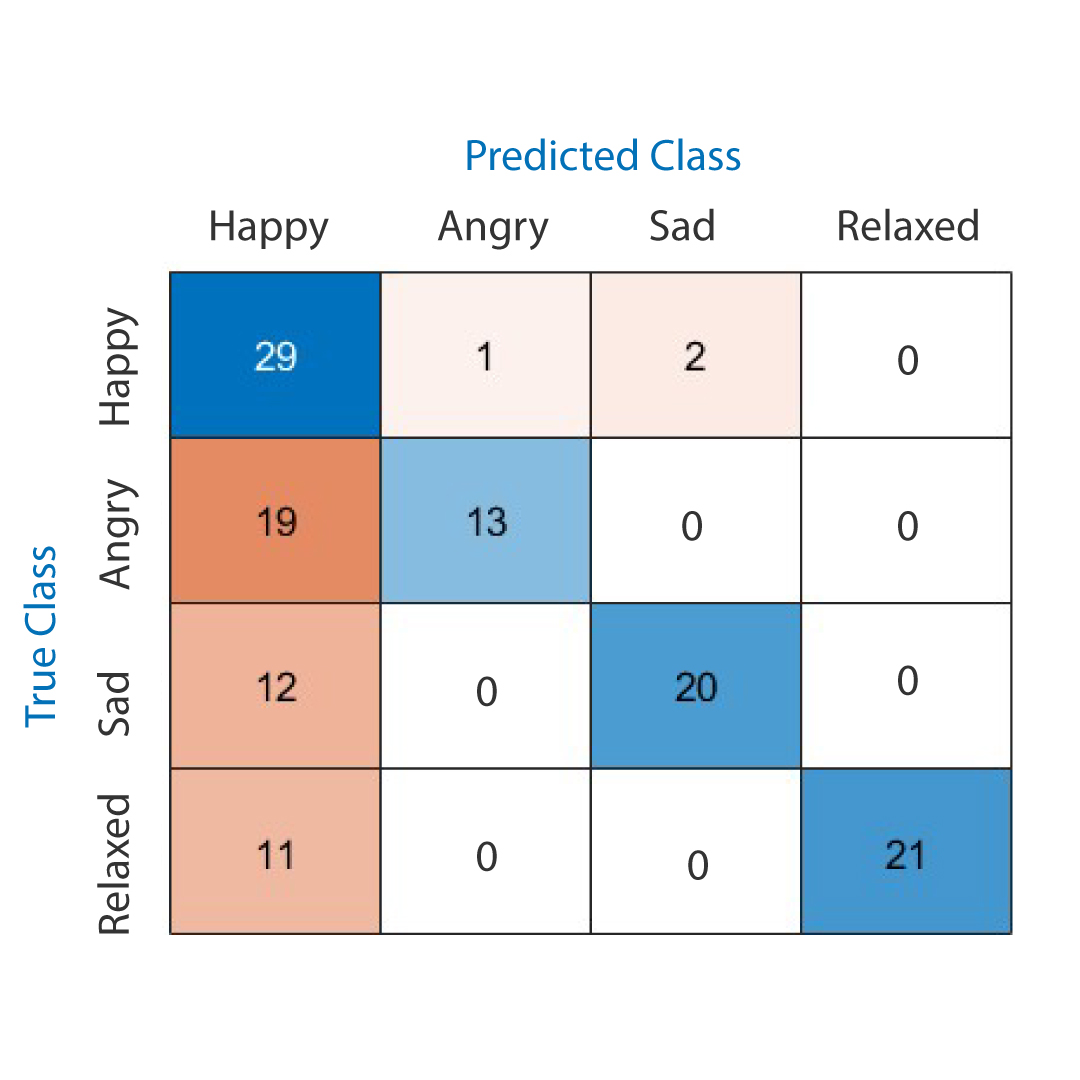}
        \caption{}
        \label{fig:subfig_c}
    \end{subfigure}
    \hfill
    \begin{subfigure}{0.23\textwidth}
        \includegraphics[width=\linewidth]{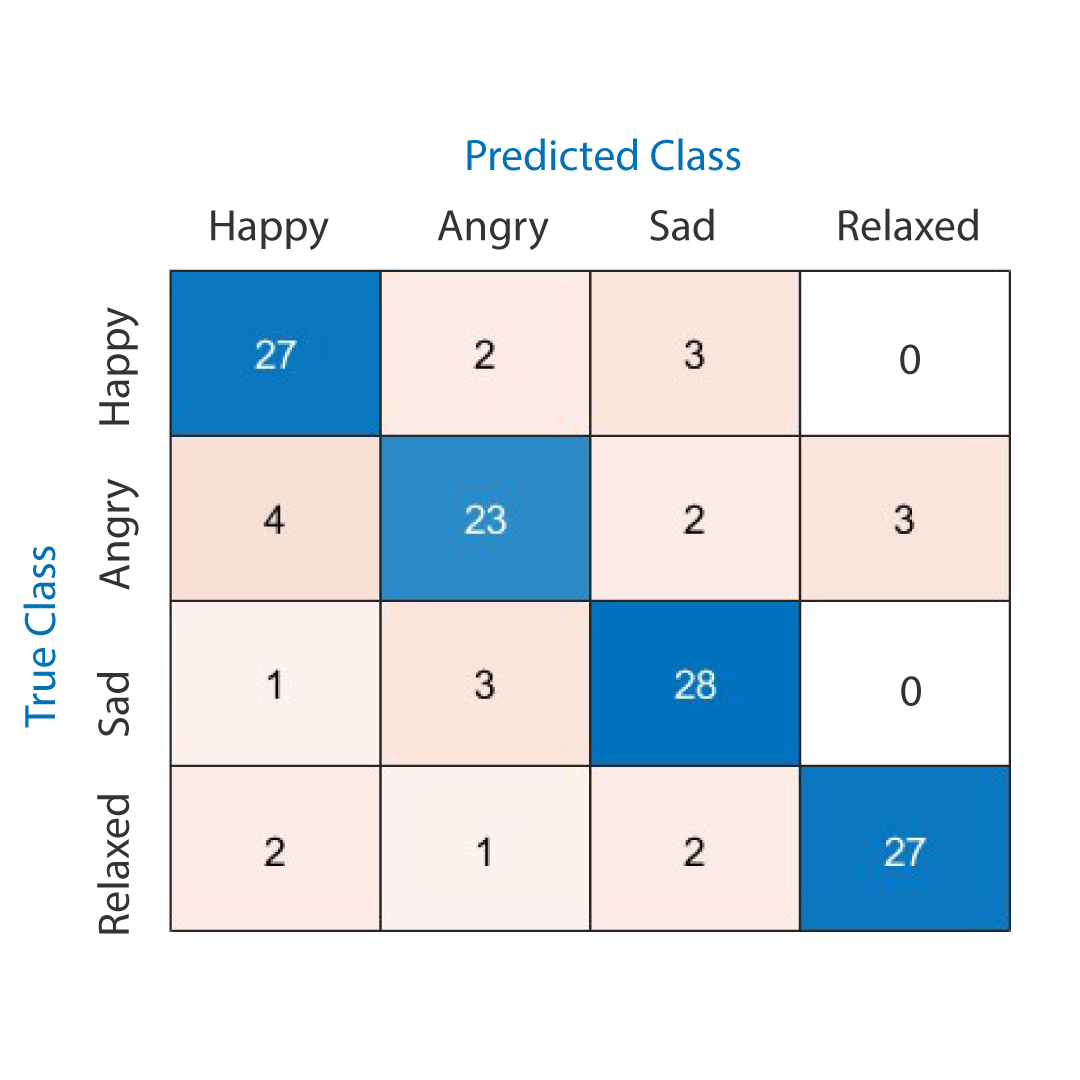}
        \caption{}
        \label{fig:subfig_d}
    \end{subfigure}
    
    \caption{Confusion matrix on the test set for (a) RBF, (b) linear, (c) Gaussian, (d) polynomial. kernels}
    \label{fig: overall}
\end{figure}

Fig. 2.  shows the confusion matrix for all four kernels used for 4-class classification on the test set. In evaluating the performance of the four kernel functions (RBF, Linear, Gaussian, and Polynomial) on a four-class classification problem (Happy, Angry, Sad, Relaxed), distinct strengths and weaknesses emerge. The Polynomial kernel exhibits the highest average precision 0.82, recall 0.81, and F1 score of 0.82 across all classes, showcasing its superior performance. On the other hand, the Linear kernel demonstrates balanced precision across classes. The RBF and Gaussian kernels perform similarly, excelling in distinguishing instances in the "Happy" and "Relaxed" classes but facing challenges in accurately classifying instances in the "Angry" and "Sad" categories.

Table II presents the performance comparison of our proposed method with recent studies conducted for the classification of emotions in response to VR content. Studies are selected for comparison as they used VR 360-degree videos for emotion elicitation stimuli. The comparison was made based on the number of participants, the number of EEG channels used, and the type of classification. A study conducted in [15], that used VR stimuli to elicit the target emotions. EEG and ECG signals were recorded for 60 participants. Features were extracted from the recorded signals using principal component analysis (PCA). Extracted features were then trained using SVM to recognize the desired emotions. Leave-One-Subject-Out (LOSO) cross-validation method was used. The model achieved an accuracy of 75.0\% for arousal and 71.0\% accuracy for the valence dimension. In [17], a VR-based EEG dataset named DER-VREED  was presented. 25 participants (15 males and 10 females) were engaged in the experiment. 60 3D videos were used each 4 sec long, and 20 videos for each positive, negative, and neutral emotion. A 64-channel wireless EEG device was used to collect the signals. The dataset is currently in the embargo phase and is not yet available publicly. Another VR-based EEG dataset presented in [18] that targets four classes of emotions namely happy, scared, calm, and bored. 32 individuals participated and watched 39 3D VR videos.

\begin{table}
    \centering

    \caption{Performance comparison of our proposed method with some existing studies that used VR content in emotion recognition using EEG signals.}
    \label{table_example}
    \centering
    \begin{tabular}{|>{\centering\arraybackslash}m{1cm}|>{\centering\arraybackslash}m{1.25cm}|>{\centering\arraybackslash}m{1cm}|>{\centering\arraybackslash}m{1cm}|>{\centering\arraybackslash}m{1cm}|>{\centering\arraybackslash}m{0.75cm}|}
        \hline
        Method & Subjects & EEG channels & Accuracy & Classifier & Type \\
        \hline 
        DER-VREEG [17] & 32 & 4 & 85.01\% & SVM & 4 class \\
        \hline
        VREED [18] & 25 & 60 & 71.35\% & SVM & 3 class \\
        \hline
        [15] & 60 & 10-20 & 75\% & SVM & Based on Valance and Arousal \\
        \hline
        Proposed & 40 & 4 & \textbf{85.54\%} & SVM & 4 class \\
        \hline
    \end{tabular}
\end{table}

\section{CONCLUSION}

In this paper, we used VR 360-degree videos as emotion elicitation stimuli for collecting an EEG dataset, for human emotions analysis and recognition. From raw EEG signals, we extracted five feature groups named correlation, power spectral density, power spectrum, rational asymmetry, and divisional asymmetry. These features were then used to train SVM with different kernels to classify four classes of emotions. Results confirmed that the polynomial kernel outperformed other kernels with a maximum average accuracy of  85.54\% on 10-fold cross-validation and 82.03\% accuracy on testing.  This paper not only bridges a gap in VR-based emotion datasets but also establishes a foundation for the integration of emotion recognition into future VR technologies. In the future, we intend to extend our work to multimodal studies and use other physiological signals as well along with EEG recordings.

\addtolength{\textheight}{-12cm}   





\begin{thebibliography}{99}
\bibitem{c1} Joy, E., Joseph, R. B., Lakshmi, M. B., Joseph, W., \& Rajeswari, M. “Recent survey on emotion recognition using physiological signals.” 7th International Conference on Advanced Computing and Communication Systems (ICACCS), pp. 1858-1863, 2021.

\bibitem{c2} S. Alarcao and M. Fonseca, "Emotions Recognition Using EEG Signals: A Survey", IEEE Transactions on Affective Computing, vol. 10, no. 3, pp. 374-393, 2019. 

\bibitem{c3} P. Samal and R. Singla, ”EEG Based Stress Level Detection During Gameplay,” 2021 2nd Global Conference for Advancement in Technology (GCAT), pp. 1-4, 2021.

\bibitem{c4} S. Mantri, D. Patil, P. Agrawal, and V. Wadhai, ”Non-invasive EEG signal processing framework for real-time depression analysis", 2015 SAI Intelligent Systems Conference (IntelliSys), pp. 518-521, 2015.


\bibitem{c5} Sun, J., Tang, Y., Lim, K.O., Wang, J., Tong, S., Li, H. and He, B., ”Abnormal Dynamics of EEG Oscillations in Schizophrenia Patients on Multiple Time Scales,” in IEEE Transactions on Biomedical Engineering, vol. 61, no. 6, pp. 1756-1764, 2014. 

\bibitem{c6}A. Ptukhin, K. Serkov, A. Khrushkov and E. Bozhko, "Prospects and modern technologies in the development of VR/AR," 2018 Ural Symposium on Biomedical Engineering, Radioelectronics and Information Technology (USBEREIT), Yekaterinburg, Russia, pp. 169-173, 2018.

\bibitem{c7} Dwivedi, Y.K., Hughes, L., Baabdullah, A.M., Ribeiro-Navarrete, S., Giannakis, M., Al-Debei, M.M., Dennehy, D., Metri, B., Buhalis, D., Cheung, C.M. and Conboy, K., "Metaverse beyond the hype: Multidisciplinary perspectives on emerging challenges, opportunities, and agenda for research, practice and policy", International Journal of Information Management, vol. 66, p.102542, 2022.


\bibitem{c8} Koelstra, S., Muhl, C., Soleymani, M., Lee, J.S., Yazdani, A., Ebrahimi, T., Pun, T., Nijholt, A. and Patras, I., "DEAP: A Database for Emotion Analysis; Using Physiological Signals", IEEE Transactions on Affective Computing, vol. 3, no. 1, pp. 18-31, 2012.

\bibitem{c9} S. Katsigiannis and N. Ramzan, "DREAMER: A Database for Emotion Recognition Through EEG and ECG Signals from Wireless Low-cost Off-the-Shelf Devices", IEEE Journal of Biomedical and Health Informatics, vol. 22,no. 1, pp. 98-107, 2018.

\bibitem{c10} R. Subramanian, J. Wache, M. Abadi, R. Vieriu, S. Winkler and N. Sebe,"ASCERTAIN: Emotion and Personality Recognition Using Commercial Sensors", IEEE Transactions on Affective Computing, vol. 9, no. 2, pp. 147-160, 2018.

\bibitem{c11} D. Liao et al., "Design and Evaluation of Affective Virtual Reality System Based on Multimodal Physiological Signals and Self-Assessment Manikin", IEEE Journal of Electromagnetics, RF, and Microwaves in Medicine and Biology, vol. 4, no. 3, pp. 16-224, 2020.

\bibitem{c12} Kim, A., Chang, M., Choi, Y., Jeon, S. and Lee, K., "The effect of immersion on emotional responses to film viewing in a virtual environment", In 2018 IEEE Conference on Virtual Reality and 3D User Interfaces (VR),pp. 601-602, 2018.

\bibitem{c13} B. Meuleman and D. Rudrauf, "Induction and Profiling of Strong Multi-Componential Emotions in Virtual Reality", IEEE Transactions on Affective Computing, vol. 12, no. 1, pp. 189-202, 2021.

\bibitem{c14} Liao, D., Shu, L., Liang, G., Li, Y., Zhang, Y., Zhang, W. and Xu, X., "Design and evaluation of affective virtual reality system based on multimodal physiological signals and self-assessment manikin", IEEE Journal of Electromagnetics, RF and Microwaves in Medicine and Biology, vol. 4, no. 3, pp. 216-224, 2019.

\bibitem{c15} Marín-Morales, J., Higuera-Trujillo, J.L., Greco, A., Guixeres, J., Llinares, C., Scilingo, E.P., Alcañiz, M. and Valenza, G., "Affective computing in virtual reality: emotion recognition from brain and heartbeat dynamics using wearable sensors", Scientific reports, vol. 8, no. 1, p.13657, 2018.

\bibitem{c16} Yu, M., Xiao, S., Hua, M., Wang, H., Chen, X., Tian, F. and Li, Y., "EEG-based emotion recognition in an immersive virtual reality environment: From local activity to brain network features", Biomedical Signal Processing and Control, vol. 72, p.103349, 2022.

\bibitem{c17} N. Suhaimi, J. Mountstephens and J. Teo, "A Dataset for Emotion Recognition Using Virtual Reality and EEG (DER-VREEG): Emotional State Classification Using Low-Cost Wearable VR-EEG Headsets", Big Data and Cognitive
Computing, vol. 6, no. 1, p. 16, 2022.

\bibitem{c18} T. Xu, R. Yin, L. Shu and X. Xu, "Emotion Recognition Using Frontal EEG in VR Affective Scenes," 2019 IEEE MTT-S International Microwave Biomedical Conference (IMBioC), Nanjing, China, 2019, pp. 1-4.

\bibitem{c19} Li, B.J., Bailenson, J.N., Pines, A., Greenleaf, W.J. and Williams, L.M., "A public database of immersive VR videos with corresponding ratings of arousal, valence, and correlations between head movements and self report measures", Frontiers in psychology, vol. 8, p.2116, 2017. 




\end{thebibliography}
\end{document}